\documentclass[lettersize,journal]{IEEEtran}
\usepackage{amsmath,amsfonts}
\usepackage{algorithmic}
\usepackage{algorithm}
\usepackage{array}
\usepackage{textcomp}
\usepackage{stfloats}
\usepackage{url}
\usepackage{verbatim}
\usepackage{graphicx}
\usepackage{bm}
\usepackage{booktabs,adjustbox,makecell,multirow}
\usepackage{hyperref}
\usepackage{mathrsfs}
 \usepackage{float}
\usepackage{marvosym}
 \usepackage[justification=centering]{caption}
 \usepackage{utfsym}
\usepackage{graphicx}
\usepackage{balance}
\usepackage{cite}

\begin{document}

\title{Quality-Aware End-to-End Audio-Visual Neural Speaker Diarization}

\author{Mao-Kui He, Jun Du,~\IEEEmembership{Senior Member,~IEEE}, Shu-Tong Niu, Qing-Feng Liu, and Chin-Hui Lee,~\IEEEmembership{Fellow,~IEEE}\thanks{M.~He, J.~Du and S.~Niu are with the National Engineering Research Center of Speech and Language Information Processing, University of Science and Technology of China, Hefei, Anhui, P. R. China (e-mail: hmk1754@mail.ustc.edu.cn; jundu@ustc.edu.cn; niust@mail.ustc.edu.cn). Q. F.~Liu is with iFlytek Hefei, Anhui, P.R.China (e-mail: qfliu@iflytek.com). C.-H. Lee is with the School of Electrical and Computer Engineering, Georgia Institute of Technology, Atlanta, GA, USA (e-mail: chl@ece.gatech.edu).}}

\maketitle

\begin{abstract}

In this paper, we propose a quality-aware end-to-end audio-visual neural speaker diarization framework, which comprises three key techniques. First, our audio-visual model takes both audio and visual features as inputs, utilizing a series of binary classification output layers to simultaneously identify the activities of all speakers. This end-to-end framework is meticulously designed to effectively handle situations of overlapping speech, providing accurate discrimination between speech and non-speech segments through the utilization of multi-modal information. Next, we employ a quality-aware audio-visual fusion structure to address signal quality issues for both audio degradations, such as noise, reverberation and other distortions, and video degradations, such as occlusions, off-screen speakers, or unreliable detection. Finally, a cross attention mechanism applied to multi-speaker embedding empowers the network to handle scenarios with varying numbers of speakers. Our experimental results, obtained from various data sets, demonstrate the robustness of our proposed techniques in diverse acoustic environments. Even in scenarios with severely degraded video quality, our system attains performance levels comparable to the best available audio-visual systems.

\end{abstract}

\begin{IEEEkeywords}
speaker diarization, multi-modality, audio-visual, visual voice activity detection
\end{IEEEkeywords}

\section{Introduction}
Speaker diarization is a prominent topic within the realm of speech signal processing, focusing on the automatic identification and segmentation of individual speakers in audio and video recordings. This rapidly evolving area plays a pivotal role in a wide range of applications, including speech recognition, transcription, multimedia content indexing, speaker-based content retrieval, meeting analysis, and more. By unraveling the identities and temporal distribution of speakers in a given audio-visual stream, speaker diarization facilitates a deeper understanding of conversational dynamics, enhances user experience, and enables intelligent systems to interact seamlessly with human-generated content.

A prevalence of multimedia content, propelled by the proliferation of digital media platforms, has highlighted a necessity for effective approaches to managing audio-visual data. Conventional methods of manually annotating speakers, being labor-intensive and impractical for large-scale sets, renders automatic speaker diarization as a vital solution. These systems utilize advanced technologies from machine learning, signal processing, and pattern recognition to segment audio-visual streams into distinct speaker chucks, each representing a unique individual engaged in discussions or presentations.

To promote audio-visual speaker diarization (AVSD) research, a few audio-visual speaker diarization corpora have been released to cover many scenarios. The AMI \cite{kraaij2005ami} data set includes audio and video recordings of meetings with multiple participants. AVDIAR \cite{gebru2017audio} contains video recordings of TV shows and meetings along with synchronized audio and visual data. The AVA-AVD \cite{xu2022ava} data set provides annotations for speaker diarization labels on the AVA movie data set \cite{roth2020ava}. Both VoxConverse \cite{chung2020spot} and MSDWild \cite{liu22t_interspeech} are collected from public videos, covering a wide range of situations and languages. The MISP challenge series \cite{chen2022audio, wang2023multimodal} releases a large-scale Chinese audio and video database which is widely used for audio-visual speaker diarization and speech recognition research. These data sets offer diverse data for AVSD, including background noise and music, indoor and outdoor environments, and multilingual content. Moreover, visual modality missing is a common phenomenon for in-the-wild videos, including the partially or completely off-screen speakers, such as the speaker on the phone or radio, and voice-over.

While a majority of research on speaker diarization has traditionally focused on the audio stream alone \cite{LANDINI2022101254, ryant2019second, ryant21_interspeech, wang2018speaker, xiao2021microsoft}, recent initiatives have begun to explore the integration of both audio and visual signals \cite{noulas2011multimodal, gebru2017audio, chung2019said, xu2022ava}. These approaches either merge similarity scores from two separate modalities or capitalize on the synchrony between speech and lip movements. Nevertheless, they encounter several challenges when processing low-quality videos. The congruence between facial visuals and vocal characteristics serves as a valuable indicator for speaker diarization. For instance, a female speaker's voice might typically have a higher frequency, while a heavier individual might exhibit a broader range of sound intensities. Naturally, humans are adept at noticing these cross-modal associations, enabling them to distinguish and categorize speakers in extended videos. However, the prevalent late fusion technique does not fully capitalize on the potential connections between speech and facial expressions. Some studies have explored speaker enrollment based on audio-visual correspondences, yet this approach fails to enroll speakers who are not visually present in the video frames, reducing its efficacy in cases involving off-screen speakers.

This work represents an extension of a previously proposed end-to-end audio-visual neural speaker diarization framework \cite{he2022end}, which contains three key improvements. Firstly, our audio-visual model takes audio features, multi-speaker visual inputs, and multi-speaker speaker embedding as multimodal inputs, utilizing a series of binary classification output layers to simultaneously identify the activities of all speakers. We describe our network framework in detail in the following aspects. The inputs consist of audio and visual features (e.g., lip regions of interest (ROI)) and speaker embedding (e.g., i-vectors and x-vectors). Besides, we have also replaced the previously employed LSTM with the transformer-based decoder.
Secondly, we utilize a quality-aware audio-visual fusion structure to effectively mitigate signal quality issues in both audio (e.g., noise, music, reverberation, or other distortions) and video (e.g., occlusions, off-screen speakers, low resolution or unreliable detection). It is generally believed that when there is a problem with the quality of a certain modality, the audio and video are out of sync. So we train the audio and visual feature extraction network by contrastive learning as in \cite{chung2017out}. The extracted audio-visual feature pairs provide the confidence of sync. We incorporate the confidence into the audio-visual cross attention \cite{tao2021someone} to automatically choose different kinds of fusion strategies including full fusion, partial fusion and not fusion. Experiments show that the quality-aware fusion strategy well handle the problems existed in both audio and video modalities, especially the lip missing and low resolution.
Lastly, to facilitate the handling of varying numbers of speakers more conveniently, we introduced a cross-speaker attention mechanism as a replacement for directly concatenating multi-speaker features, which typically requires setting a maximum number of speakers that the network can handle. Thanks to the cross-speaker mechanism, the new network can seamlessly adapt to situations with any number of speakers, eliminating the need for preset limitations.

\section{Related Work}
\textbf{Audio-only speaker diarization} has garnered extensive research attention. Conventionally, speaker diarization systems comprise distinct, independent sub-modules. The initial step typically involves voice or speech activity detection (VAD/SAD), which is responsible for distinguishing speech from non-speech events \cite{johnston2012webrtc}. Following this, raw speech signals within the identified speech segments undergo transformation into acoustic features or embedding vectors \cite{dehak2010front,snyder2018x,wan2018generalized}. In the subsequent clustering stage, these transformed speech segments are grouped and assigned labels corresponding to speaker classes \cite{von2007tutorial,park2019auto}. Lastly, in the post-processing stage, the results obtained from clustering are subjected to further refinement. Each of these sub-modules is optimized individually in general. Recently, the methods that incorporate deep learning into entire speaker diarization systems has been more and more popular. They are optimized in a multi-speaker situation to learn the relations between speakers and output multi-speaker speech activity probabilities. The target-speaker voice activity detection (TS-VAD) \cite{medennikov2020target} is proposed to achieve accurate speaker diarization even under noisy conditions with many speaker overlaps. TS-VAD takes not only a sequence of MFCC, but also a set of i-vectors or x-vectors \cite{cheng2023target} as inputs. The model outputs a sequence of the speech activity probabilities of all speakers simultaneously. Because TS-VAD requires the i-vectors of speakers, pre-processing to obtain the i-vectors is necessary, which also is taken as post-processing of clustering-based diarization. The fully end-to-end neural diarization (EEND) \cite{fujita2019end}  performed all the speaker diarization procedures based on a single neural network. EEND was initially proposed using a bidirectional long short-term memory (BLSTM) network, and was soon extended to the self-attention-based network \cite{fujita2019end}. Extension of EEND with the encoder–decoder-based attractor (EEND-EDA) \cite{horiguchi2022encoder} applies an LSTM-based encoder–decoder on the output of EEND to generate multiple attractors. Attractors are generated until the attractor existing probability becomes less than the threshold. Then, each attractor is multiplied by the embedding generated from EEND to calculate the speech activity for each speaker. Although both TS-VAD and EEND based methods have improved diarization performance compared with clustering-based systems, practical applications are limited by generalization ability on unseen domains. Those audio-based methods still struggle to diarize the low-quality (heavy noise and reverberation), high overlapping speech.

\textbf{Audio-visual speaker diarization} emerges as a solution to overcome the limitations of audio-only methods. Facial attributes and lip motion are closely related to speech activity \cite{yehia1998quantitative}, suggesting a synergy between utterances and visual cues. Over the last decade, methods utilizing mutual information, canonical correlation analysis, and deep learning have sought to exploit this relationship. Early efforts were largely based on the Bayesian framework \cite{noulas2011multimodal, gebru2017audio, 10096295}, with dynamic bayesian networks modeling individuals in audio-visual recordings as multimodal entities. Researchers have combined visual tracking with speech-source localization to address speech-to-person association \cite{gebru2017audio}, introducing audio-visual fusion techniques for mapping acoustic features to visual data and determining speaker turns through latent-variable temporal models. Recent advancements include a temporal audio-visual fusion model \cite{10096295} for diarization, notable for its computational efficiency and robustness without the need for training. This model tracks dominant speakers by aligning sound locations with visual presence.
Sound source localization, employing circular or linear microphone arrays, provides a basis for another cross-modal interaction by projecting sound direction onto the visual plane, enhancing speaker localization accuracy. \cite{cabanas2018multimodal} proposed a system using a modified SRP-PHAT function evaluated on space volumes to localize speakers based on microphone arrays.

Audio-visual synchronization ensures the alignment of audio and visual elements in multimedia presentations, such as videos or films, achieving precise timing between sound and corresponding visuals for viewer coherence. Techniques, including the two-stream ConvNet \cite{krizhevsky2012imagenet} architecture \cite{chung2017out}, have been developed to learn synchronization from unlabeled data, significantly enhancing the audio-visual speaker diarization process. This process, crucial for identifying and tracking speakers in multimedia recordings, benefits immensely from accurate synchronization \cite{chung2019said,ding2020self}. Furthermore, clustering-based methods in audio-visual systems have advanced, with proposals like the audio-visual relation network (AVR-Net) \cite{xu2022ava} that assesses and scores the similarity between audio and visual speaker features. This facilitates the generation of diarization results through clustering based on these similarity scores. Additionally, the dynamic vision-guided speaker embedding (DyViSE) \cite{wuerkaixi2022dyvise} method utilizes dynamic lip movement and facial features to improve audio clarity and establish distinct speaker identities for each segment, showcasing the evolving landscape of speaker diarization technologies.

\section{Quality-Aware End-to-End Audio-Visual Speaker Diarization framework}
\label{AVSD_framework}

\begin{figure*}[t]
  \centering
  \includegraphics[width=\linewidth]{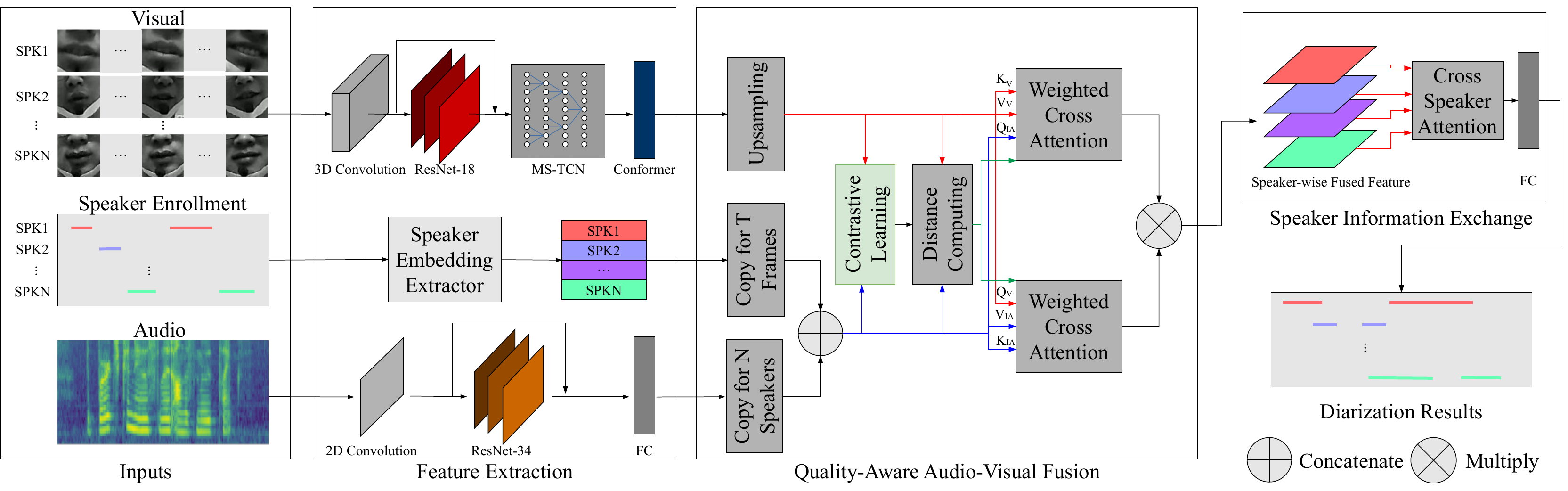}
  \caption{The illustration of network structure}
  \label{fig:avsd_framework}
\end{figure*}

Figure~\ref{fig:avsd_framework} illustrates the proposed quality-aware end-to-end audio-visual speaker diarization network. Moving from left to right, the comprehensive inputs encompass visual elements such as faces or lips, speaker enrollment, and audio spectrum characteristics. This is succeeded by a feature extraction block designed to extract visual features, speaker embedding, and audio features. The quality-aware audio-visual fusion initially assesses whether the audio and visual components are synchronized and then autonomously selects the fusion strategies. Ultimately, the speaker information exchange phase processes the fused audio-visual feature and calculates the frame-level speech/silence probabilities for each speaker. In this section, we will first introduce the representation of the entire framework under a probabilistic model, then provide a detailed description of the audio-visual speaker diarization network framework, and finally present the optimization process.

\subsection{Probabilistic Model}
We use a probabilistic model to decompose the proposed AVSD framework into several parts, each of which is modeled with different neural networks. To handle all possible cases in our model, let $N$ be the number of speakers in a video including off-screen talkers. We set the undetected or off-screen speakers' lip ROIs to all zeros, then at time $t$ we have $N$ lip ROIs $\textbf{\textup{X}}_t=(\emph{\textbf{X}}_{t}^1,...,\emph{\textbf{X}}_{t}^n,...,\emph{\textbf{X}}_{t}^N) \in \mathbb{R}^{N \times W \times H}$. The observed random variable $\emph{\textbf{X}}_{t}^n \in \mathbb{R}^{W\times H}$ is the visual feature with width $W$ and height $H$ of person $n$ at time $t$. For the audio-based data, without loss of generality, we use $\textbf{\textup{Y}}_t=(\emph{Y}_{t,1},...,\emph{Y}_{t, f},...,\emph{Y}_{t, F}) \in \mathbb{R}^{F}$ to denote the $F$-dimensional FBANKs of single-channel audio signals. The time series $\textbf{\textup{X}}_{1:T}=\{\textbf{\textup{X}}_1,...,\textbf{\textup{X}}_t,...,\textbf{\textup{X}}_T\}$ and $\textbf{\textup{Y}}_{1:T}=\{\textbf{\textup{Y}}_1,...,\textbf{\textup{Y}}_t,...,\textbf{\textup{Y}}_T\}$ represent the visual and audio observations, respectively. The objective of speaker diarization is to assign speech signal to persons. For this purpose, we introduce a time series of discrete variables $\textbf{\textup{S}}_{1:T}=\{\emph{\textbf{S}}_1,...,\emph{\textbf{S}}_t,...,\emph{\textbf{S}}_T\} \in \{0,1\}^{N\times T}$ where the vector $\emph{\textbf{S}}_t=(S_{t}^1,...,S_{t}^n,...,S_{t}^N) \in \{0,1\}^N$ has binary-valued elements so that $S_{t}^n=1$ if speaker $n$ is speaking during the time-step $t$, and $S_{t}^n=0$ if speaker $n$ is not speaking. For brevity, we ignore the subscript $1:T$ hereinafter.

Obviously, when visual streams show silence caused by visual modality missing or unreliable, previous models \cite{noulas2011multimodal,gebru2017audio} are confused about which speaker to assign for current speech. To avoid the alignment problem, we introduce speaker embedding $\textbf{\textup{I}}=\{\emph{\textbf{I}}^1,...,\emph{\textbf{I}}^n,...,\emph{\textbf{I}}^{N} \} \in \mathbb{R}^{N \times D_\text{I}}$ as speaker related observations where $\emph{\textbf{I}}^n \in \mathbb{R}^{D_\text{I}}$ is the representation of speaker $n$. The speaker embedding is estimated with speaker replantation models on target speaker speech segments. Then the temporal speaker diarization problem can be formulated as finding the most probable time series of state $\hat{\textbf{\textup{S}}}$ among all possible speaker label sequences $\mathcal{S}$ with the observed variables $(\textbf{\textup{X}},\textbf{\textup{Y}},\textbf{\textup{I}})$, as follows:
\begin{equation}
\hat{\mathbf{S}}= \mathop{\textup{arg max}}_{\mathbf{S} \in \mathcal{S}} P(\mathbf{S}|\textbf{\textup{X}},\textbf{\textup{Y}},\textbf{\textup{I}})
\end{equation}
where $P(\mathbf{S}|\textbf{\textup{X}},\textbf{\textup{Y}},\textbf{\textup{I}})$ can be factorized using conditional independence assumption as follows:
\begin{equation}
P(\mathbf{S}|\textbf{\textup{X}},\textbf{\textup{Y}},\textbf{\textup{I}}) = P(\mathbf{S} | \mathbf{E}_\text{V},\mathbf{E}_\text{A},\mathbf{I}) P(\mathbf{E}_\text{V}|\mathbf{X}) P(\mathbf{E}_\text{A}|\mathbf{Y})
\end{equation}
where $\mathbf{E}_\text{V}=\{\emph{\textbf{E}}_\text{V}^1,...,\emph{\textbf{E}}_\text{V}^n,...,\emph{\textbf{E}}_\text{V}^N\} \in \mathbb{R}^{N \times T \times D_\text{V}}$ is the frame-level $D_\text{V}$-dimensional visual latent variables for $N$ speakers, namely V-embedding. $\mathbf{E}_\text{A} \in  \mathbb{R}^{T \times D_\text{A}}$ is the frame-level $D_\text{A}$-dimensional audio latent variables, namely A-embedding. Here, we assume the V-embedding and A-embedding are conditioned independently on the visual and audio observations respectively. Both A-embedding and V-embedding are extracted by their respective feature networks.

\begin{figure*}[t]
  \centering
  \includegraphics[width=\linewidth]{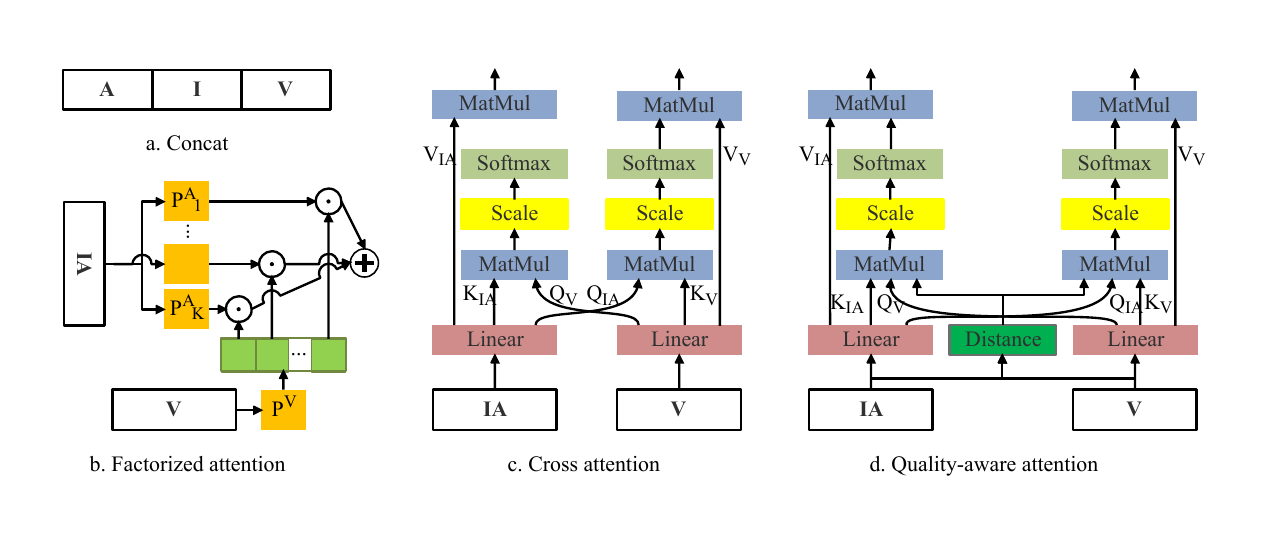}
  \vspace{-1cm}
  \caption{The four fusion strategies of audio embedding (\textbf{A}), video embedding (\textbf{V}) and speaker embedding (\textbf{I}). \textbf{IA} are concatenation of audio and speaker embedding.}
  \label{fig:fusion}
\end{figure*}

\subsection{Quality-Aware Audio-Visual Speaker Diarization Networks}

\subsubsection{Feature Extraction}
\label{sec:feature}
We build a visual network to compute V-embedding of the $n$-th person $\emph{\textbf{E}}_\text{V}^n \in \mathbb{R}^{T \times D_\text{V}}$ by directly adding several conformer \cite{gulati2020conformer} layers to the original multiscale temporal convolutional networks (MS-TCN) for lipreading \cite{martinez2020lipreading}. The visual network is illustrated in the top row of Figure~\ref{fig:avsd_framework}. By projecting the V-embedding $\emph{\textbf{E}}_\text{V}^n$ to the frame-level speech/non-speech probabilities $\hat{\emph{\textbf{S}}}_\text{V}^n=(\hat{S}_\text{V}^n(1),...,\hat{S}_\text{V}^n(t),...,\hat{S}_\text{V}^n(T)) \in [0,1]^T$ for the $n$-th person through full connected (FC) layers, the whole network can be used as visual voice activity detection (V-VAD). Meanwhile, the visual-only speaker diarization results can be directly generated with V-VAD by combining VAD of all persons in one session. However, these results perform very poorly with off-screen speakers or missing visual frames, where visual features cannot provide any information of speech activities. Moreover, audio and visual modalities do not align exactly. Lip wiggling while not speaking leads to false alarm and some words are not accompanied by lip movements. Therefore, there is almost no diarization research specifically focused on pure video. In our system, the visual-only results, together with the audio-only results, serve as initial segmentation for extracting speaker embedding.

Since most scenes are accompanied by reverberation, we first perform audio dereverberation with NARA-WPE \cite{drude2018nara} as in \cite{he2022ustc}. FBANKs are taken as inputs to an audio feature extractor. In our previous work \cite{he2022end}, the extractor consists of a CNN with 2D Convolution, BatchNorm, and Relu per layer. Here, we replace the extractor with ResNet-34 \cite{he2016deep} as in \cite{cheng2023target} and finally use 2D Convolution to make the number of input and output frames the same. Finally, a fully connected layer projects high dimensional features to low dimensional A-embedding $\mathbf{E}_{A}$. Unlike visual networks, we do not pre-train the audio network separately on any other tasks and optimize it with the entire model. This network is illustrated in the bottom row of Figure~\ref{fig:avsd_framework}.

There are two kinds of assignment problems without speaker embedding. When a speaker's lips are not visible, the output may either not label any speech or randomly assign the speech label to an individual. When one person is speaking and another is not, yet both are moving their lips, the output may label both individuals or randomly select one as the speaker. Those two errors occur in the cases of occlusions, off-screen speakers, error detection and moving lips, although someone is not speaking. In audio-only speaker diarization approaches \cite{tsvad_Medennikov}, multi-speaker voice activities are estimated simultaneously by taking both audio features and their speaker embedding as inputs. Therefore, we introduce speaker embedding (S-embedding), including i-vector \cite{dehak2010front} and x-vector \cite{desplanques20_interspeech}, to deal with the unreliable V-embedding problem in audio-visual diarization mentioned above. In the training stage, we compute the audio speaker embedding with non-overlapping segments of each speaker with oracle labels. In the inference stage, we estimate audio speaker embedding with audio-only and visual-only speaker diarization results. This process is depicted on the left side of the middle row in Figure~\ref{fig:avsd_framework}.

\subsubsection{Quality-Aware Audio-Visual Fusion}
\label{sec:QAFusion}

Audio-visual fusion aims to leverage the complementary nature of audio and visual cues, improving the overall performance of AVSD. The key point related to feature fusion is to make good use of complementary information to improve system robustness. Here, we first introduce three existing fusion networks, and then propose a new strategy, quality-aware audio-visual fusion, to address the problem of incomplete modality accuracy. The four fusion structures are illustrated in Figure~\ref{fig:fusion}.

In earlier work, we concatenated the audio, visual, and speaker embedding along the time axis and fed them into a shared speaker detection (SD) component, which comprises a 2-layer bidirectional LSTM with projection (BLSTMP). While this straightforward feature fusion strategy handles normal visual inputs effectively, it experiences a significant performance decline when visual inputs are incomplete due to occlusions, off-screen speakers, unreliable detection, or low resolution.
Factorized attention \cite{li2023audio} is a method used in the audio-visual speech separation module to integrate audio and visual embedding effectively. In factorized attention, the acoustic  and visual embedding are first transformed separately using linear matrices and then factorized into multiple acoustic subspace vectors, while the visual embedding is projected onto a lower-dimensional vector. These transformed embedding vectors are then fused together using a sigmoid function to obtain the final audio-visual embedding.
Cross attention is a mechanism used in TalkNet \cite{tao2021someone}, an audio-visual active speaker detection (ASD) model, to learn the interacted new audio feature. In audio stream, visual feature is used as the target sequence to generate the query, and audio feature is used as the source sequence to generate the key and value. In visual stream, the query is generated with audio feature, and the key and value is generated with visual feature. This allows the model to dynamically align the audio and visual content and learn the audio-visual relationship. The cross attention layer is followed by a feed-forward layer, and residual connection and layer normalization are applied to generate the whole cross-modal attention network. The outputs of the cross attention network are concatenated along the temporal direction.

To address the quality unstable problem in audio and visual modalities, we proposed a new fusion strategy named quality-aware audio-visual fusion. The entire fusion process can be split into two stages including audio-visual synchronization and weighted cross attention. In the first stage, the network plays a role of audio-visual synchronization \cite{chung2017out} to give a confidence in whether the current audio-visual pair is synchronized. Given an unseen audio-visual clip, the modal quality problem can be transformed into a synchronization problem, assuming that at least one modal quality is not problematic (If both modalities are untrustworthy, correct results cannot be obtained from the current signal). Synchronization means that both the audio and visual inputs are of high quality. On the contrary, at least one modality is of poor quality. Unlike traditional audio-visual synchronization task where the audio and visual pairs usually have only one speaker, overlapping speech is often present in speaker diarization tasks. Here, we concatenate speaker and audio embedding to generate speaker-wise audio embedding $\mathbf{E}_\text{IA}  \in \mathbb{R}^{N \times T \times D_\text{IA}}$. With the contrastive learning on the visual and speaker-wise audio embedding, the entire feature extraction network realizes the audio-visual synchronization under overlapping segments. For stability, we compute the average L2 distance $\mathcal{L}^n_{t}$ of speaker $n$ between the visual and speaker-wise audio embedding at frame $t$ on a small window $T_w$:
\begin{equation}
  \mathcal{L}^n_{t}=\frac{1}{2T_w+1} \sum_{\tau=t-T_w}^{t+T_w}\lVert \emph{\textbf{E}}_\text{IA}^n(\tau)-\emph{\textbf{E}}_\text{V}^n(\tau) \rVert
\end{equation}
where $ \emph{\textbf{E}}_\text{IA}^n(\tau) \in \mathbb{R}^{D_\text{IA}}$ and $\emph{\textbf{E}}_\text{V}^n(\tau)  \in \mathbb{R}^{D_\text{V}}$ are the speaker-wise audio and visual embedding of speaker $n$ at frame $\tau$. The L2 distance is normalized through the following operation:
\begin{equation}
  W_{t}^n=\frac{M}{M+\mathcal{L}_t^n}
\end{equation}
where the weight $W_{t}^n$ is used for subsequent audio-visual feature fusion. $M$ is a constant term used to control the numerical range of $W_{t}^n$. In our experiments, we set $M=1$ and the range of $W_{t}^n$ is normalized to $(0,1]$. For speaker $n$, we denote the weight sequence as $\emph{\textbf{W}}^n=\{W_1^n,...,W_t^n,...,W_T^n\} \in \mathbb{R}^T$.

In the second stage, we design a soft audio-visual fusion including fully and constrained cross attention between audio and visual features, and self-attention on audio and visual feature separately. The inputs of speaker $n$ are the vectors of query $(\textbf{Q}_\text{IA}^n,\textbf{Q}_\text{V}^n)$, key $(\textbf{K}_\text{IA}^n,\textbf{K}_\text{V}^n)$, and value $(\textbf{V}_\text{IA}^n,\textbf{V}_\text{V}^n)$ from speaker-wise audio and visual embedding, respectively, projected by a linear layer. The outputs are the audio attention feature $\textbf{F}_{\text{IA} \to \text{V}}^n$ and visual attention feature $\textbf{F}_{\text{V} \to \text{IA}}^n$ as formulated in Eqs.~(\ref{eq:qa1}) and (\ref{eq:qa2}) as shown below,

\begin{equation}
\label{eq:qa1}
 \textbf{F}_{\text{IA} \to \text{V}}^n=\\
 s \left( \frac{((\emph{\textbf{W}}^n)^\top\textbf{Q}_\text{V}^n + (1-\emph{\textbf{W}}^n)^\top\textbf{Q}_\text{IA}^n)(\textbf{K}_\text{IA}^n)^\top}{\sqrt{d}} \right) \textbf{V}_\text{IA}^n
\end{equation}

\begin{equation}
\label{eq:qa2}
 \textbf{F}_{\text{V} \to \text{IA}}^n=\\
 s \left( \frac{((\emph{\textbf{W}}^n)^\top\textbf{Q}_\text{IA}^n + (1-\emph{\textbf{W}}^n)^\top\textbf{Q}_\text{V}^n)(\textbf{K}_\text{V}^n)^\top}{\sqrt{d}} \right) \textbf{V}_\text{V}^n
\end{equation}
where $s$ is softmax. As formulated in Eqs.~(\ref{eq:qa1}) and ~(\ref{eq:qa2}), to learn the interacted new audio feature $\textbf{F}_{\text{IA} \to \text{V}}^n$, the attention layer applies $(\emph{\textbf{W}}^n)^\top\textbf{Q}_\text{V}^n + (1-\emph{\textbf{W}}^n)^\top\textbf{Q}_\text{IA}^n$ as query, and $\textbf{E}_\text{IA}^n$ as the source sequence to generate key $\textbf{K}_\text{IA}^n$ and value $\textbf{V}_\text{IA}^n$, and to learn $ \textbf{F}_{\text{V} \to \text{IA}}^n$, vice versa. The new defined query performs different fusion strategies under different weights. When $\emph{\textbf{W}}^n$ approaches 1, the audio and visual modalities are well synchronized and a full cross-attention will occur. When $\emph{\textbf{W}}^n$ is close to 0, the audio and visual features are out of sync and will perform self-attention. When $\emph{\textbf{W}}^n$ ranges from 0 to 1,  a constrained cross attention will be performed.
The attention layer is followed by a feed-forward layer. Residual connection and layer normalization are also applied after these two layers to generate the whole cross-modal attention network. The outputs are taken as the new speaker-wise audio and visual embedding and fed into the next quality-aware fusion layer. The speaker-wise audio and visual embedding of the last layer are multiplied in the temporal direction to obtain the speaker-wise fused feature $ \textbf{E}_\text{AV}^n$ for speaker $n$.

\subsubsection{Speaker Information Exchange}

\begin{figure}[t]
  \centering
  \includegraphics[width=\linewidth]{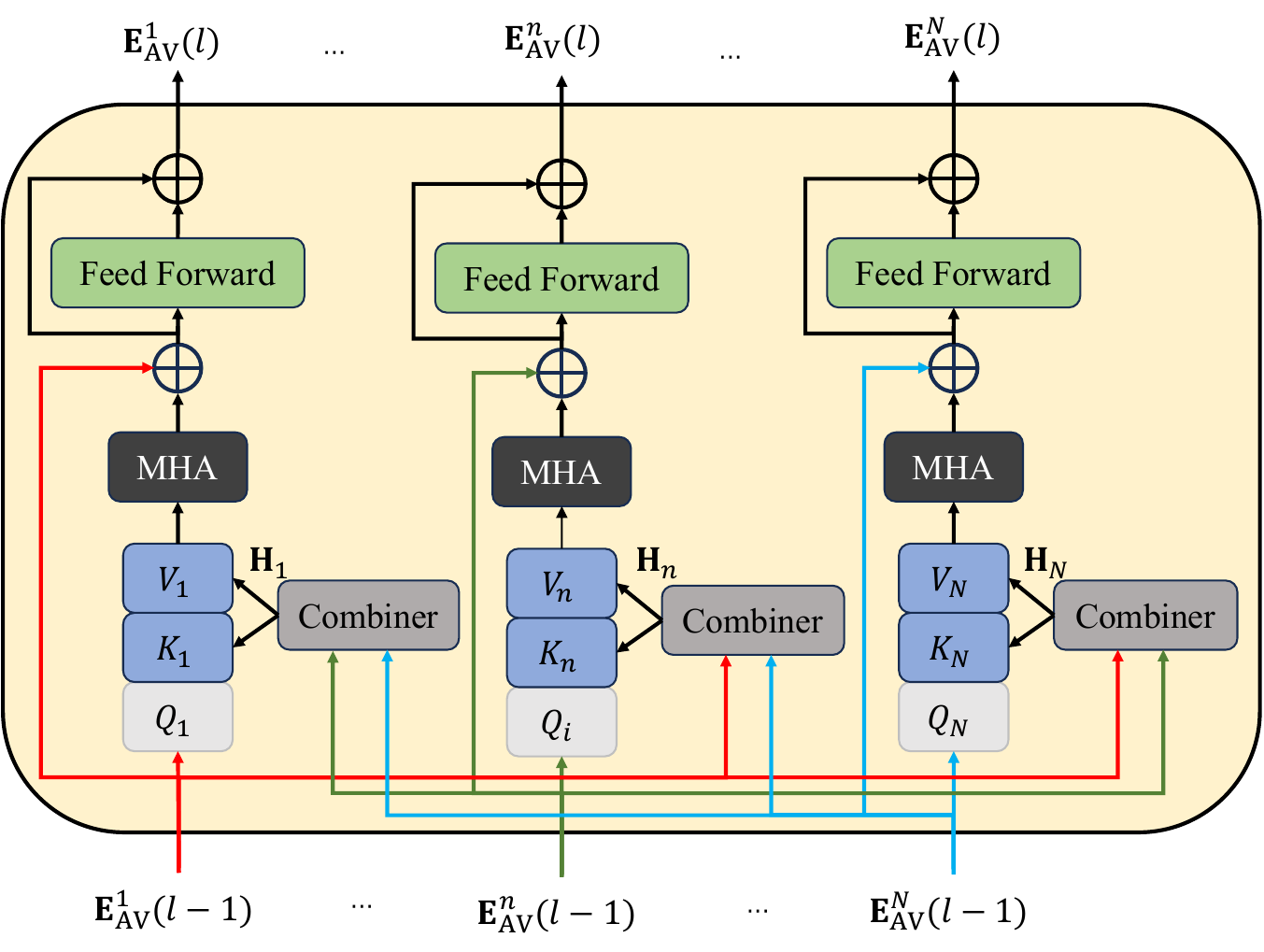}
  \caption{Illustration of the cross-speaker attention layer. `MHA' stands for multi-head attention.}
  \label{fig:cross_attention}
\end{figure}

One of the main differences from audio-visual active speaker detection is that audio-visual speaker diarization needs to specify which person is speaking. Therefore, information exchange between speakers is very important, which benefits the model in distinguishing speaker roles and overlapping speech segments. The speaker-wise embedding is concatenated directly in earlier audio-only studies (e.g., TS-VAD). Our previous AVSD  \cite{he2022end} work followed this setup, concatenating speaker-wise fused feature along the time series. However, this approach requires setting in advance the maximum number of speakers the network can handle, thus limiting usage scenarios. If the maximum number of speakers is set too small, only the scenes with very few people can be handled. If the maximum number of speakers is increased, although it can handle scenes with more people, it will increase the unnecessary calculation amount when dealing with scenes with a small number of people. To address this problem, we adopt the cross-speaker attention structure to improve speaker interactions. The cross-speaker attention layer is similar to the cross-channel attain layer proposed in \cite{chang2021multi}, where we replace channel interaction with speaker interaction and prove that it works under various scenarios with different number of speakers. As shown in Figure~\ref{fig:cross_attention}, given the $N$-speaker audio-visual fusion features $\mathbf{E}_\text{AV}$, these attention layers aim to learn the contextual relationship across speakerz both within and across time steps. When we use the speaker-wise fused features of speaker $n$ to create $\mathbf{Q}_\text{AV}^n$, the other speakers are leveraged by a combiner to create $\mathbf{K}_\text{AV}^n$ and $\mathbf{V}_\text{AV}^n$. Here, we only investigate an combiner that takes the average of the fused features of other speakers,
\begin{equation}
\textbf{H}_n=\frac{1}{N-1} \sum_{i \neq n}^N{\mathbf{E}_\text{AV}^i}
\end{equation}
which can be seen as the symmetric weight case of the Affine combiner. Unlike in \cite{chang2021multi}, we don't take the other combiner which concatenates the other channels along the time axis, because this will limit the model capability for handling variable speaker numbers. Each cross-speaker attention layer can process all of speakers through parameter sharing.
With this adaptation, the model parameters do not increase w.r.t. the number of speaker ($N$) and time frames ($T$). Furthermore, the cross-speaker attention outputs of the audio-visual fusion features at the $l^\text{th}$ layer $\mathbf{E}_\text{AV}(l) \in \mathbb{R}^{N \times T \times D_\text{AV}}$, where $D_\text{AV}$ is the feature dimension and $\mathbf{E}_\text{AV}(0)$ is the speaker-wise fused feature vector  $\mathbf{E}_\text{AV}$ discussed in Section~\ref{sec:QAFusion} and shown in Figure ~\ref{fig:cross_attention}.

Finally, we take a shared linear layer to map audio-visual fusion features of each speaker to $N$ outputs $\hat{\textbf{S}}=\{\hat{\textbf{S}}^{1},...,\hat{\textbf{S}}^{n},...,\hat{\textbf{S}}^{N}\} \in (0,1)^{N \times T}$ corresponding to the speech/non-speech probabilities for each of the $N$ speakers respectively. The speaker information exchange block is illustrated in the upper-right side in Figure~\ref{fig:avsd_framework}.

\subsection{Model Optimization}
\label{sec:Optimizing}
The proposed audio-visual speaker diarization network is optimized in the following three steps:
First, we copy the parameters of the pre-trained lipreading model to the visual network and initialize the audio network randomly. The training objective is that the outputs of the audio and video feature extraction networks are similar for genuine pairs, and different for false pairs. Specifically, the Euclidean distance between the network outputs is minimized or maximized. We use the contrastive loss \cite{chung2017out}:
\begin{equation}
\label{eq:cl}
J_\text{C}=\frac{1}{T}\sum_{t=1}^{T}z_t\mathcal{L}_t^2+(1-z_t)[\max(margin-\mathcal{L}_t,0)]^2
\end{equation}
where $z_t \in [0, 1]$ is the binary similarity metric between the audio and the video inputs. Genuine pairs include not only single-person speech clips \cite{chung2017out} but also overlapping speech. Ensure that fully cross-modal fusion can be achieved as long as the overlapping speech parts are synchronized, because the visual modality can greatly help the speaker diarization on the overlapping segments \cite{he2022end}.

Next, we freeze the audio and video feature extraction networks and train a speaker information exchange block as illustrated in the upper-right side in Figure~\ref{fig:avsd_framework} to predict better diarization results. The loss function is expressed as follows:
\begin{equation}
J_\text{AV}=\frac{1}{NT}\sum_{n=1}^{N}\sum_{t=1}^{T}{BCE(S^n_t, \hat{S}^n_t)}
\end{equation}
where $BCE$ is the binary cross entropy function between the labels and the outputs.

Finally, we unfreeze the audio and video feature extraction networks and train the whole network jointly with a small learning rate. Contrastive loss is kept to retain the synchronization characteristics of the audio and video feature extraction network. The loss function is written as follows:
\begin{equation}
J=\lambda \cdot J_\text{C}+ J_\text{AV}
\end{equation}
where $\lambda = 0.1$ in our experiments.

We also build a visual VAD network to obtain initial speaker enrollment mentioned in section~\ref{sec:feature}. The visual embedding from the trained visual feature extractor is taken as inputs and mapped to frame-level speech/non-speech probability sequences using LSTM layers and fully connected layers. The visual feature extractor is always frozen. The visual embedding is available on both V-VAD and AVSD networks, which reduces redundant computations.

\section{Experiment and Result Analysis}

\subsection{Audio-Visual Speaker Diarization Corpus}
Currently, various audio-visual diarization corpora have been released. The AMI corpus \cite{kraaij2005ami} is a multi-modal dataset consisting of 100-hour meeting recordings. The close-talking and far-field audio, individual and room-view video are all available. Since each person is equipped with a close-up camera, the recorded video signal is of relatively high quality. AMI is an ideal data set for speaker diarization. AVDIAR \cite{gebru2017audio} is released to enable audio-visual scene analysis of unstructured informal meetings and gatherings, but it does not provide the training set. VoxConverse \cite{chung2020spot} is a challenging dataset where the diverse speaker diarization data from `in the wild' videos are included. However, the video signal of VoxConverse is not available now. AVA-AVD \cite{xu2022ava} has been released recently to cover diverse video characteristics and complex acoustic conditions of movie scenes, including speakers completely off-screen. Recently, the newly released multi-modal information based speech processing (MISP) dataset \cite{chen2022audio, wang2023multimodal} provides more than 100-hour audio and video recordings of several people in a living room watching and chatting while interacting with a smart speaker/TV. These sessions are usually accompanied by high overlap ratios in multi-talker conversions and real domestic noise backgrounds such as TV, air conditioning and movements. MSDWild \cite{liu22t_interspeech} is collected from public videos and covers a wide range of scenarios and languages. A comparison of those data sets and their corresponding features is shown in Table~\ref{tab:dataset_comparsion}.

Most of the previous studies provided experimental results on AMI, our ablation experiments were also conducted on AMI. We evaluate our model following \cite{LANDINI2022101254} to create training, development, and test sets where only words are considered as speech. We evaluate Mix-Headset audio for the ablation study and microphone array audio for comparison with findings from other studies. To demonstrate the robustness of the proposed method, we also compare our best model with existing systems on other publicly available audio-visual speaker diarization datasets, including AVA-AVD, MSDWild, and MISP2021 \cite{chen2022audio}. We meticulously adhere to the guidelines for partitioning these datasets into training, development, and testing sets.

\begin{table*}
\centering
\setlength{\abovecaptionskip}{10pt}
\setlength{\belowcaptionskip}{10pt}
\caption{A comparison of six existing audio-visual speaker diarization data sets in terms of four features in separate columns: (1) \textbf{speech} for speech amount in videos, (2) \textbf{Overlapped} for overlapped speech rate per video, (3) \textbf{\#speakers} for the min/average/max speaker number per video, (4) \textbf{Noise} for whether the videos contain noises (+) or not (-). The `Access' column indicates each corpus with access to audio (A+) and video (V+) or without video (V-).}
\label{tab:dataset_comparsion}
\begin{adjustbox}{max width=\linewidth}
\begin{tabular}{lccccccccc}
\toprule
Datasets & Access & Scenario & \#Videos & Duration & Speech (in \%) & Overlapped (in \%) & \#Speakers & Noise  & Language \\
\midrule
AMI \cite{kraaij2005ami} & A+/V+ &Meetings & 170 & 100h & 80.9 & 13.6 & 3/4.0/5  &  - &EN\\
AVDIAR \cite{gebru2017audio} & A+/V+ &Chat & 27 & 21m & 82.6& 10.5  & 1/2.2/4 & - &EN, FR\\
VoxConverse \cite{chung2020spot} & A+/V- &TV show  & 448   & 63h50m & 90.7 & 3.6   & 1/5.6/21 & - &EN\\
AVA-AVD \cite{xu2022ava} & A+/V+ &Movie & 351 & 29h15m & 46.0 &  4.4  & 2/7.7/24  & + &Multi\\
MSDWild \cite{liu22t_interspeech} &A+/V+ &Daily conversation  & 3143 & 80h3m & 91.3&  14.0& 2/2.7/10 & + &Multi\\
MISP \cite{chen2022audio, wang2023multimodal} & A+V+ &Home conversation & 373 & 120h53m & 92.3 & 25.8 & 2/4.2/6   & + &CN\\
\bottomrule
\end{tabular}
\end{adjustbox}
\end{table*}

\subsection{Evaluation Metric}
The accuracy of speaker diarization system in this track is measured by diarization error rate (DER) \cite{fiscus2006rich} which is calculated as the summed time durations of three different errors of false alarm (FA), missed detection (MISS) and speaker errors (SPKERR) divided by the total time duration. All of these components are the lower the better. We report DER with 0s collar in most cases, which means no forgiveness collar will be applied to the reference segments prior to scoring and overlapping speech will be evaluated. In some AVSD articles, only the results with a `collar=0.25s' setting are provided, and for the sake of facilitating comparison, we also report these corresponding results based on their settings. The oracle VAD was provided to fix diarization results in our previous work. To improve the practicality of the proposed AVSD model, we don't use any reference information to correct the model results in this paper.

\subsection{Human Face Detection and Tracking}
Due to the diversity of audio-visual data sources, audio sample rate and video frame rate are not exactly the same, especially data coming from the Internet. To facilitate unified processing of models, we standardize the audio sampling rate for all datasets to 16 kHz and the video frame rate to 25 frames per second (fps).

Some datasets (e.g., MISP2021) provide the locations of face and lip ROIs, we don't repeatedly extract those visual features. Most datasets only provide raw videos without these facial position information. Therefore, we perform human face detection and tracking on those video data based on the mainstream techniques that have been open-sourced so far. Specifically, we employ RetinaFace \cite{Deng2020CVPR} to extract all face images and the corresponding 5 facial landmarks in each video. We use Deep SORT \cite{Wojke2017simple} for face tracking and take Arcface \cite{deng2019arcface} as appearance descriptor. Next, we employ the Kalman filter algorithm to predict the state of the tracks from the previous frame in the current frame. Then, we compute the cost matrix between the predicted tracks and the detections in the current frame using the Mahalanobis distance computed based on the appearance information, and perform cascade matching and IOU matching on them to obtain all matched track pairs, unmatched tracks, and unmatched detections in the current frame. For each case, we process them separately: for each successfully matched track, we update it with the detection corresponding to the current frame; for the unmatched track, we delete it; for the unmatched detection, we initialize them as a new track. At the same time, we add facial landmark recognition to track lip positions. The above preprocessing operations enable us to obtain face and lip sequence of the video.

\subsection{Speaker Enrollment}

In our previous work, speaker enrollment is completely obtained from the V-VAD network (visual-only results) because there are almost no speakers with complete off-screen in MISP2021. But for real-world scenarios, this is not an exception, which causes off-screen speakers to be lost in the results. Here, we use the fusion of audio-only and visual-only results to include as many speakers as possible. For audio-only system, we employ a state-of-the-art clustering-based method, namely, VBHMM x-vector diarization \cite{LANDINI2022101254} (i.e., VBx). The x-vectors are first extracted with the deep neural network architecture based on ResNet101 \cite{zeinali2019but} for each speech segment divided by VAD which is predicted by time delay neural network (TDNN) \cite{ryant21_interspeech}. Then, they are clustered by means of AHC with a similarity metric based on probabilistic linear discriminant analysis (PLDA) log-likelihood ratio scores \cite{prince2007probabilistic}, followed by VB-HMM-based clustering to create the diarization results.

\subsection{Network Configuration}
We extracted 40-dimensional FBANKs (i.e., $F=40$) with 25 ms frame length and 10 ms frame shift as audio feature inputs. The detected lip or face inputs were resized to width 96 pixels and height 96 pixels ($W \times H=96 \times 96$). Since the video is re-sampled to 25 fps (40 ms frame shift), the 256-dimensional ($D_\text{V}$) V-embedding was repeated $4$ times for each frame to align a single audio frame. The speaker embedding extractors, including 100-dimensional ($D_\text{I}$) i-vector and 256-dimensional x-vector based on ECAPA-TDNN \cite{desplanques20_interspeech} were pre-trained on VoxCeleb1 \cite{Nagrani19}, VoxCeleb2  \cite{Chung18b} and CN-CELEB \cite{fan2020cn}. We will show all the speaker embedding, results in section \ref{Network_results}. In our earlier work, A-embedding is extracted with a 4-layer 2D CNN. We replace the simple CNN with ResNet-34 \cite{he2016deep}. The 256-dimensional ($D_\text{A}$) A-embedding and speaker embedding are concatenated and projected to 256-dimensional ($D_\text{IA}$) speaker-wise audio embedding. We used three quality-aware attention blocks for audio-visual feature fusion where each block was equipped with 256 encoder dims, 4 attention heads, 32 convolution kernel size for both audio and visual streams. We conducted four types audio-visual fusion structures as mentioned in section \ref{sec:QAFusion} to show the robustness of the proposed quality-aware fusion strategy. Since there were only audio and visual features for fusion in factorized attention and cross attention, we simply concatenate audio and speaker embedding in our framework as the speaker-related audio features. With this setting, we can directly integrate factorized and cross attention into our AVSD framework for feature fusion. To facilitate interaction between speaker-wise fused feature streams, we introduce cross-speaker attention layers, each equipped with 256 encoder dimensions, 4 attention heads, and a convolution kernel size of 32. The impact of the number of attention layers on the results will be discussed in section~\ref{sec:cross_speaker}.

\subsection{Network Training and Inference}

The proposed AVSD network training process is divided into three stages, including contrastive learning of audio and visual feature extractors, separate training of speaker information exchange blocks, and joint training of the entire AVSD network. For the contrastive learning stage, we obtained genuine (positive) and false lip-sync audio-video training samples by cropping the AVSD datasets. And those clips where speakers are speaking in overlapping segments will also be regarded as genuine pairs. In order to improve the generalization ability of the model, we adopt the following data augmentation strategies. For audio stream, we utilize the NARA-WPE for dereverberation. And input audio is augmented with 3-fold speed perturbation randomly. Besides, MUSAN \cite{snyder2015musan} and RIRs \cite{7953152} are applied as the additional audio augmentation. For visual stream, input videos undergo each item of the following procedures with a probability of 0.5: rotation, horizontal flipping; cropping with the scale range, transformation of contrast, brightness, and saturation. We randomly pick video frames and corrupt the entire input feature, including replacing other people's lips, giving random values from 0 to 255, and setting all zeros. This greatly improves the model's robustness under conditions of very poor video quality.

After model inference, we obtain the frame-level speech/silence probability of each speaker, and then generate the final diarization results through a series of post-processing. Firstly, thresholding is applied to the probabilities associated with the speakers to produce preliminary results. Then, we merge segments with short silence intervals for each speaker to address the issue where the AVSD model may mistakenly predict brief pauses by the speaker as silence, whereas they are labeled as speech in the manual annotations. Finally, We apply DOVER-Lap to the multi-channel results if the data set includes multi-channel audio. Subsequently, we estimate speaker embedding by utilizing the AVSD results as a basis for speaker enrollment, which proves to be more reliable than audio-only or visual-only outcomes, and then repeat the above process. The optimal threshold value for each data set is determined using the development set (DEV) and subsequently applied to the evaluation set (EVAL).

\subsection{Results and Analysis}
In this section, we ﬁrstly show the results of the proposed AVSD framework including the input type of visual feature and speaker representation as mentioned in section \ref{sec:feature}. Then, we demonstrate the results of different fusion structures and analyze the effects of various visual modality problems including detection missing caused by occlusions, off-screen speakers or unreliable detection and low resolution of visual features as the person is too far away from the camera. Next, we illustrate the effectiveness of the cross-speaker attention structure compared to direct concatenation, which limits the maximum number of speakers that can be processed. Finally, we compare our best results with existing systems on publicly available audio-visual speaker diarization data sets.
\begin{table*}
    \centering
    \small
    \caption{An ablation study of the proposed AVSD framework evaluated on the AMI EVAL set showing the diarization performances of four techniques and their corresponding alternatives: (1) audio encoder with CNN or ResNet-34, (2) visual input with Lip or Face, (3) speaker embedding with i-vector or x-vector, and (4) training strategy with Direct or Joint training. The symbol `-' indicates the same configuration as `C1' while `C2' swaps the audio encoder in `C1', `C3' swaps the visual input in `C1', `C4' swaps the speaker embedding in `C1', and `C5' swaps the training strategy in `C1'}
        \begin{tabular}{ccccccccccccc}
            \toprule
            \multirow{2}{*}{Configuration} & \multicolumn{2}{c}{(1) Audio encoder} & \multicolumn{2}{c}{(2) Visual input} & \multicolumn{2}{c}{(3) Speaker embedding} & \multicolumn{2}{c}{(4) Training strategy} & \multirow{2}{*}{FA} & \multirow{2}{*}{MISS} & \multirow{2}{*}{SPKERR} & \multirow{2}{*}{DER}   \\
            \cmidrule(r{4pt}){2-3}
            \cmidrule(r{4pt}){4-5}
	  \cmidrule(r{4pt}){6-7}
	   \cmidrule(r{4pt}){8-9}
            &CNN & ResNet-34 & Lip & Face & i-vector & x-vector & Direct & Joint & & \\
	   \midrule
	    C1 & $\times$ & + & + & $\times$ & + & $\times$ &  $\times$ & + & \textbf{3.44} & 3.72 & \textbf{1.92} & \textbf{9.08} \\
           C2 &  + & $\times$ & - & - & - & - & - & - & 3.51 & \textbf{3.65} & 2.16 & 9.32 \\
	  C3 &  - & - & $\times$ & + & - & - & - & - & 4.18 & 4.81 & 2.79 & 11.78 \\
	  C4 &  - & - & - & - & $\times$ & + & - & - & 3.62 & 4.96 & 2.31 & 10.89 \\
	   C5 & - & - & - & - & - & - &  + & $\times$  & 3.59 & 4.15 & 2.88 & 10.62 \\
            \bottomrule
        \end{tabular}
    \label{tab:ablation_results}
\end{table*}
\subsubsection{Network Ablation Study}
\label{Network_results}


Table~\ref{tab:ablation_results} presents a systematic analysis of the effects of different configurations on AVSD performance, focusing on audio encoding, visual inputs, speaker embedding, and training strategies. It outlines the diarization results across five configurations (C1 through C5) evaluated on the AMI EVAL set where is the default setting is C1. Comparison between CNN (C2) and ResNet-34 (C1) shows that ResNet-34 tends to perform better, with lower False Alarm (FA), Speaker Error (SPKERR), and Diarization Error Rate (DER). However, the improvements are not markedly significant, suggesting that the complexity of ResNet-34 might not fully counterbalance the limitations imposed by the available training data. The analysis between lip (C1) and face (C3) visual inputs reveals that configurations leveraging face information incur a higher DER. This underscores the importance of lip movements in voice activity detection and speaker diarization. The integration of such visual cues remains challenging, especially in low-quality video scenarios. We also compare the use of i-vector (C1) and x-vector (C4) speaker embedding. The findings indicate that i-vector, despite being an earlier technology compared to x-vector, still offers robust performance, potentially due to its efficiency in contexts with constrained training data. The examination of different training strategies, particularly direct (C5) and joint training (C1), highlights the effectiveness of joint training in reducing DER. This suggests that multi-stage training approaches can significantly enhance the model's ability to classify and diarize speakers accurately.

\begin{figure}
  \centering
  \includegraphics[width=\linewidth]{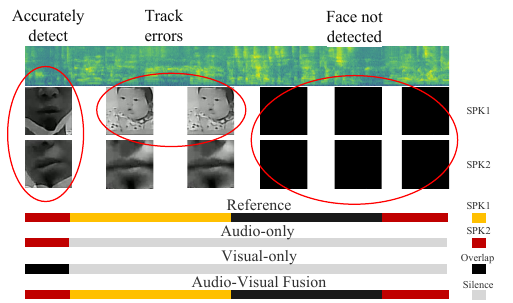}
  \caption{An example showcasing issues of track errors and missing facial detections in an audio-visual recording.}
  \label{fig:example}
\end{figure}

Figure~\ref{fig:example} depicts challenges related to track errors and the absence of facial detections. Initially, the visual-only system incorrectly classifies a speech segment as overlapping due to SPK1's lip movements. In contrast, both the audio-only and audio-visual systems accurately identify it through audio cues. Later, when SPK1 is speaking alone, both the audio-only and visual-only systems fail to detect speech, mis-attributing SPK1's lip movement to a child depicted in a wall photograph. Towards the end, the audio-visual system effectively discerns overlapping speech, despite the non-detection of lips for both SPK1 and SPK2.

\subsubsection{Audio-Visual Fusion Strategy Comparison}

\begin{table*}[t]
\centering
\caption{Lip statistics for each data set and DER of four mentioned fusion strategies across data sets.}
\label{tab:fusion_results}
\setlength{\tabcolsep}{3.5mm}{
\begin{adjustbox}{max width=\linewidth}
\begin{tabular}{clcccccc}
\toprule
&  & \multicolumn{1}{c}{\multirow{2}{*}{AMI}}  & \multicolumn{2}{c}{\multirow{1}{*}{MISP2021}}  &  \multicolumn{2}{c}{\multirow{1}{*}{MSDWild}} &\multicolumn{1}{c}{\multirow{2}{*}{ AVA-AVD}}  \\
\cmidrule(r{4pt}){4-5}
\cmidrule(r{4pt}){6-7}
 & & & Middle & Far & Few-talker & Many-talker &  \\
\midrule
\multicolumn{1}{c}{\multirow{2}{*}{Lip statistics}} & Miss rate (\%) & 11.88  & 8.11 & 14.67 & 12.78 & 23.79 & 20.33	\\
 & Resolution (pixels) & 440 & 651 & 324 & 467 & 287& 252 \\
\midrule
\multicolumn{1}{c}{\multirow{4}{*}{Fusion strategy}} & Concat & 10.56  & 10.11 & 13.68 & 11.56 & 26.89 & 23.45	\\
 & Factorized attention & 10.64 & 10.79 & 13.05 & 11.77 & 27.18 & 23.77 \\
 & Cross attention & 10.29  & 10.45  & 12.78 & 10.98 & 25.55 & 22.15 \\
 & Quality-aware fusion & 9.08 & 8.84 & 10.35 & 10.30 & 23.11 & 18.56 \\
\bottomrule
\end{tabular}
\end{adjustbox}}
\vspace{-0.5cm}
\end{table*}

Table~\ref{tab:fusion_results} presents a comparative analysis of four fusion methods applied to various AVSD datasets. The goal is to assess the performance of these methods in terms of miss rate and resolution, which are key metrics in tasks involving audio-visual processing.
The lip statistics section reveals insights into the dataset characteristics. Notably, the `Miss Rate' highlights the accuracy of lip statistic retrieval, with values ranging from 8.11\% (AMI) to 23.79\% (MSDWild Many-talker). Lip miss rate is a crucial influencing factor in audio-visual speaker diarization, reflecting the accuracy of lip information detection. Higher lip miss rates indicate less lip information provided, leading to the decreased diarization performances. 

Moreover, `Resolution' provides information on the clarity of lip images, ranging from 252 (AVA-AVD) to 651 (MISP2021 Middle) pixels. The smaller the `Resolution', the harder the AVSD is to detect.
Among the four fusion strategies, `Concat' achieves comparable performance when the miss rate is relatively low, but degrades significantly on datasets with poor video signals including higher miss rate and lower resolution.
`Factorized attention' improves when video quality is good, but performs worse under poor conditions compared with `Concat'.
`Cross attention' performs well across datasets with a lower miss rate. However, when the resolution becomes lower, the performance gain is not always significant. `Quality-aware fusion' consistently outperforms other methods in a variety of settings. This indicates that quality-aware fusion is a promising approach for accurate and clear lip statistical retrieval, benefiting from the selective fusion of modal information by weighted cross attention.

\begin{figure}[htbp]
\centering
\begin{minipage}[t]{0.5\textwidth}
\centering
\includegraphics[width=\textwidth]{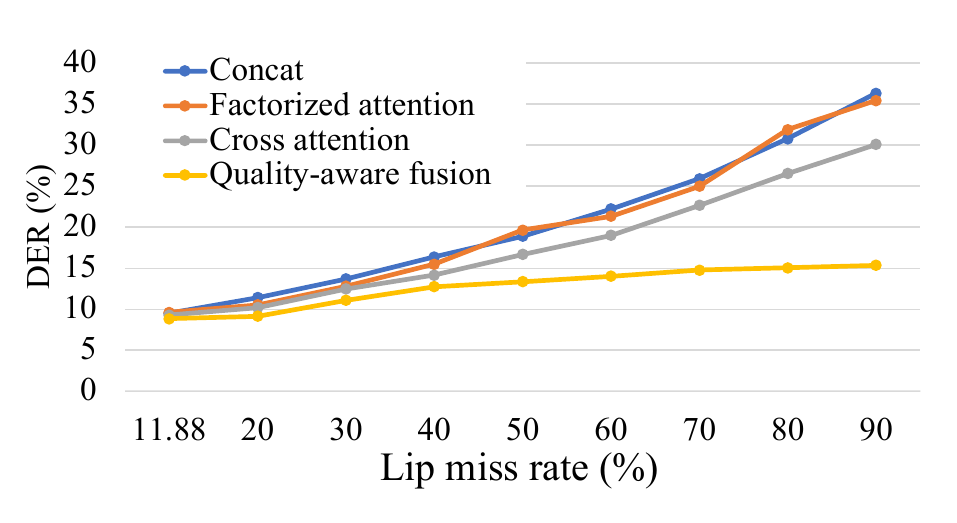}
\vspace{-0.6cm}
\caption{A DER comparison of different lip miss rates with four fusion strategies on the AMI EVAL set.}
\label{fig:LipMissRate}
\end{minipage}
\begin{minipage}[t]{0.5\textwidth}
\centering
\vspace{0.3cm}
\includegraphics[width=\textwidth]{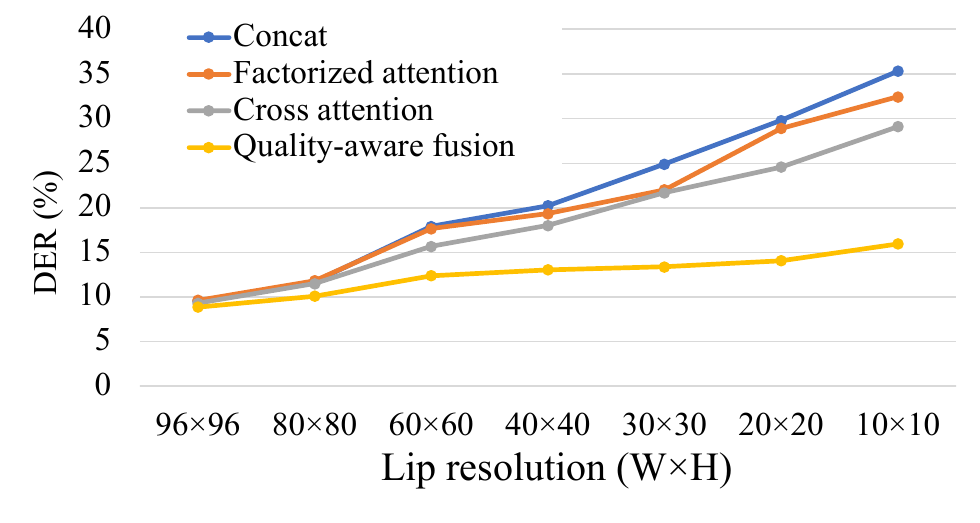}
\vspace{-0.6cm}
\caption{The DER comparison of different lip resolution with four fusion strategies on AMI EVAL set.}
\label{fig:LipResolution}
\end{minipage}
\end{figure}

To fully evaluate processing capabilities of the fusion strategies in various video environments, we artificially set lip miss rate and resolution on AMI. Figure~\ref{fig:LipMissRate} presents data on lip miss rates for different fusion strategies across various percentages from 11.88\% (Proportion of undetected lips in original data set)  to 90\%. Note that our proposed quality-aware fusion strategy is much more robust than other fusion ways. Figure~\ref{fig:LipResolution}  provides valuable insights into how lip resolution impacts diarization results under four fusion strategies. Quality-aware Attention consistently demonstrates superior performance, emphasizing its effectiveness in leveraging lip information for diarization tasks across different resolution. Factorized Attention also shows competitive performances, while Cross Attention and Concatenation exhibit higher DER values, especially for the cases at lower lip resolution. 

\subsubsection{Effects of Cross-speaker Attention Structure}

\label{sec:cross_speaker}
\begin{table}
    \centering
    \small
    \caption{A DER comparison of the proposed cross-speaker attention layers on AMI and AVA-AVD EVAL sets.}
        \begin{tabular}{lcccc}
            \toprule
            Interaction & Network & \#Layers & AMI & AVA-AVD \\
           \midrule
           Direct concat & BLSTM & 1 & 10.45 &  21.02\\
           \midrule
           Cross-   & \multirow{3}{*}{Transformer} & 3 & 10.11 & 19.28\\
           speaker &  & 4 & 9.08 & 18.56\\
            attention &  & 5 & 9.55 & 18.78 \\
            \bottomrule
        \end{tabular}
   \label{tab:cross_speaker}
\end{table}

Table~\ref{tab:cross_speaker} compares the diarization performance of different configurations or models in terms of their interaction type, network architecture, and the number of layers. Specifically, we focus on two evaluation sets: AMI and AVA-AVD. The `Direct concat' configuration with a single layer of BLSTM \cite{he2022end} achieves the highest DER on both AMI and AVA-AVD, indicating relatively poorer performance compared to other configurations. In particular, the results on AVA-AVD are far worse than the results of cross-speaker attention. This is because the number of speakers in each session of AVA-AVD  changes greatly, and the strategy of pre-setting the maximum number of people cannot handle the case of very large number of speakers. The `Cross-speaker attention' configurations using Transformer networks with multiple layers (3, 4, and 5) consistently achieve lower DER values, suggesting better diarization performance, especially in the case of 4 layers.

\subsubsection{Comparison of Results Across Data Sets}

\begin{table}[ht]
\centering
\caption{ADER comparison across data sets.}
\label{tab:sota}
\begin{adjustbox}{max width=\linewidth}
\begin{tabular}{lllc}
\toprule
 \multicolumn{2}{c}{\multirow{1}{*}{Datasets}}  & Methods & DER \\
 \midrule
 \multicolumn{1}{l}{\multirow{4}{*}{AMI}} &  \multicolumn{1}{l}{\multirow{2}{*}{Headset}} & SyncNet \cite{chung2019said} & 14.45  \\
&  & Ours & 9.08 \\
 \cmidrule{2-4}
 &  \multicolumn{1}{l}{\multirow{2}{*}{Array}} & SyncNet \cite{chung2019said} & 15.92  \\
 &  & Ours & 11.02 \\
  \midrule
  \multicolumn{1}{l}{\multirow{4}{*}{MISP2021}} &  \multicolumn{1}{l}{\multirow{2}{*}{Middle}} & MISP\_Baseline \cite{he2022end} & 10.99  \\
  &  & Ours & 8.84  \\
  \cmidrule{2-4}
  &  \multicolumn{1}{l}{\multirow{2}{*}{Far}} & MISP\_Baseline \cite{he2022end}  & 13.95  \\
  &  & Ours & 10.35 \\
  \midrule
\multicolumn{1}{l}{\multirow{4}{*}{MSDWild}} &  \multicolumn{1}{l}{\multirow{2}{*}{Few-talker}} & MSDWild\_Baseline \cite{liu22t_interspeech} & 12.20  \\
   & & Ours & 10.30 \\
 \cmidrule{2-4}
   & \multicolumn{1}{l}{\multirow{2}{*}{Many-talker}} & MSDWild\_Baseline \cite{liu22t_interspeech} & 25.86 \\
   & & Ours & 23.11 \\
   \midrule
  \multicolumn{2}{c}{\multirow{2}{*}{AVA-AVD}} & AVA-AVD \cite{xu2022ava} & 20.57  \\
  & & Ours & 18.56 \\
\bottomrule
\end{tabular}
\end{adjustbox}
\end{table}

Table~\ref{tab:sota} provides a comprehensive comparison of diarization results across different data sets and techniques. Here are the key observations:
For the AMI data set, two conditions are considered: `Headset' and `Array'.
In both conditions, our proposed framework consistently outperforms the SyncNet-based methods, achieving much lower DER values. This suggests the effectiveness of our techniques on the AMI data set.
The MISP2021 data set is divided into `Middle' and `Far' field conditions.
Our audio-visual method demonstrates superior performance with the lowest DER than the previous method under both field conditions.
The MSDWild data set is evaluated for both `Few-talker' and `Many-talker' scenarios.
In both cases, our method consistently outperforms the `MSDWild\_Baseline', indicating its effectiveness in diarization on the MSDWild \cite{liu22t_interspeech} data set.
For the AVA-AVD data set, our method achieves a lower DER than that for the`AVA-AVD' method.
In summary, our proposed approach consistently shows promising performances, achieving lower DER values when compared to other methods across different data sets and conditions. This suggests that the proposed method is effective and robust for diarization tasks in a variety of contexts.

\section{Conclusion}

In this paper, we propose an audio-visual speaker diarization framework that integrates a quality-aware fusion strategy and a cross-speaker attention mechanism, yielding promising improvements in speaker diarization performances. The quality-aware fusion strategy, which dynamically adapts the fusion method according to the synchronization state of audio and visual inputs and utilizes lip information across various resolution, consistently surpasses alternative fusion approaches in terms of robustness. Moreover, the cross-speaker attention mechanism overcomes the limitations of having to pre-define a maximum number of speakers, resulting in reduced diarization error rates, especially in environments with a high number of speakers. Looking ahead, we plan to extend the application of the quality-aware fusion approach to other audio-visual tasks, such as audio-visual speech enhancement, separation, and recognition, among others.


\bibliographystyle{IEEEtran}

\bibliography{mybib}

\vspace{0.5cm}
\begin{IEEEbiography}[{\includegraphics[width=1in, height=1.25in, clip,keepaspectratio]{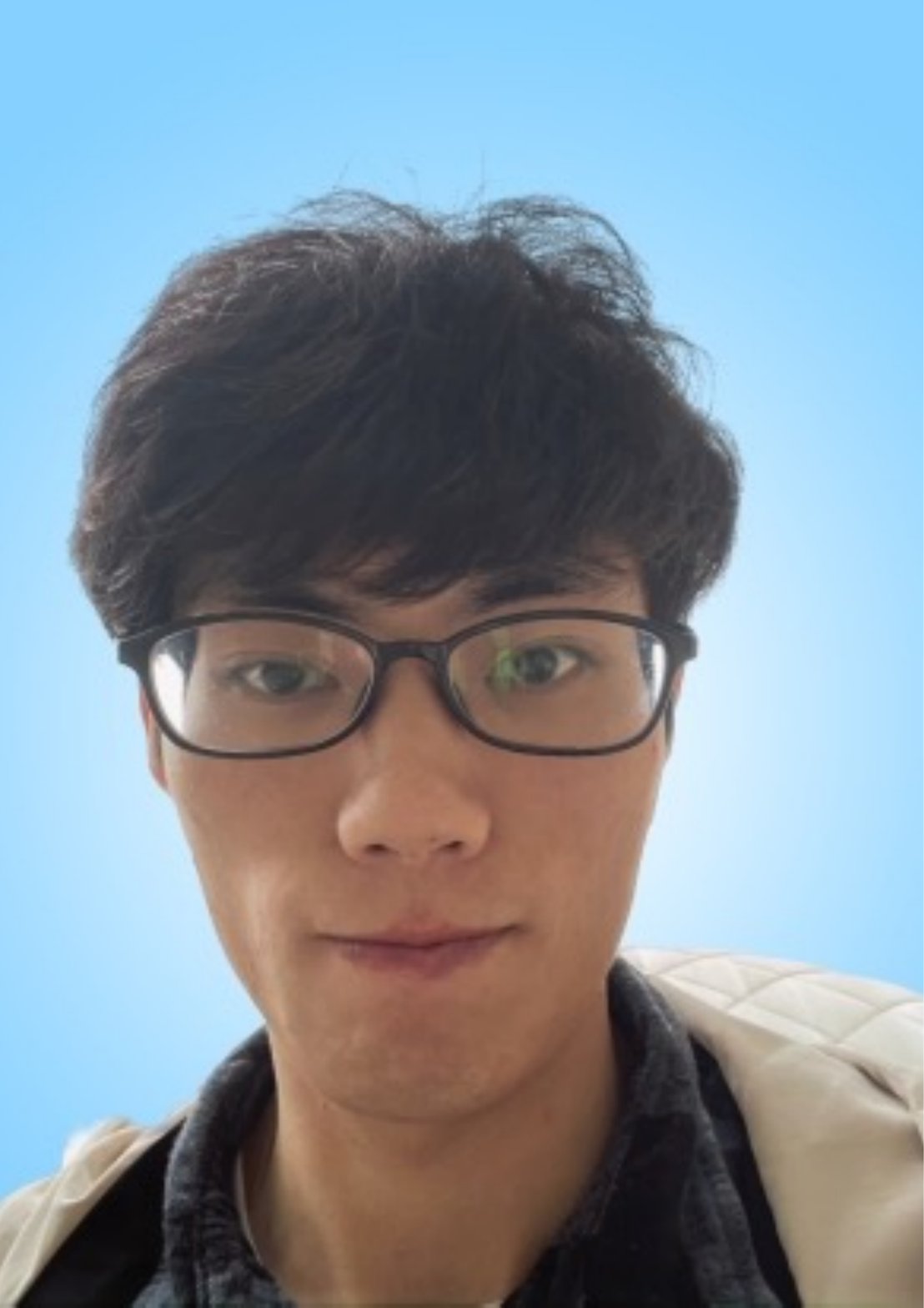}}]{Mao-Kui He}
received the B.S. degree from the Department of Electronic Engineering and Information
Science, University of Science and Technology of China (USTC), in 2018. He is currently working toward a Ph.D. degree with the National Engineering Laboratory for Speech and Language Information Processing, USTC. His recent research includes speech enhancement and speaker diarization.
\end{IEEEbiography}

\vspace{0.5cm}
\begin{IEEEbiography}[{\includegraphics[width=1.15in, height=1.25in, clip,keepaspectratio]{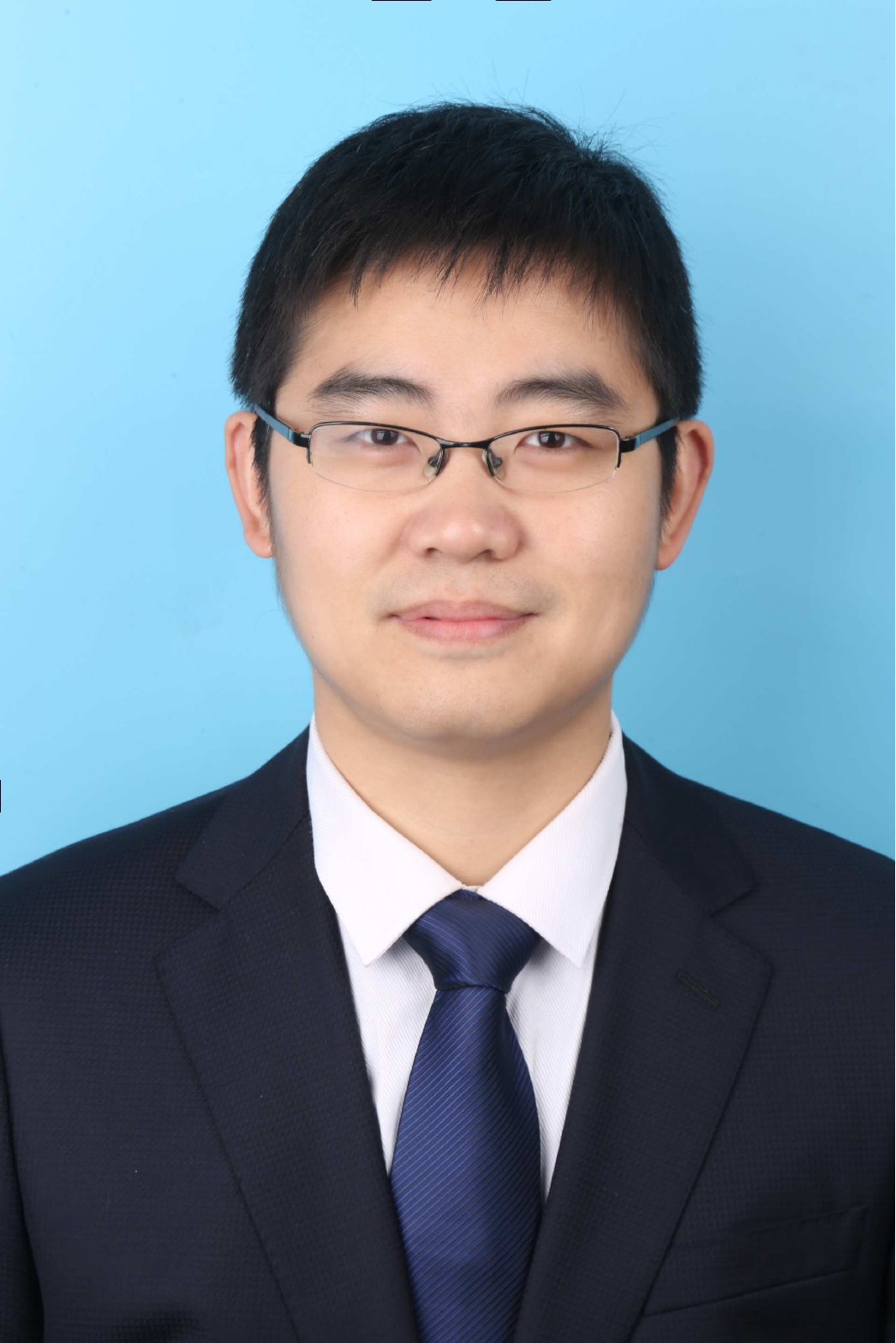}}]{Jun Du}
(Senior Member, IEEE) received a B.Eng. and Ph.D. from the Department of Electronic Engineering and Information Science, University of Science and Technology of China (USTC), Hefei, China, in 2004 and 2009, respectively. From 2009 to 2010, he was with iFlytek Research as a Team Leader, working on speech recognition. From 2010 to 2013, he joined Microsoft Research Asia as an Associate Researcher, working on handwriting recognition, OCR. Since 2013, he has been with the National Engineering Laboratory for Speech and Language Information Processing, USTC. He has authored or coauthored more than 200 papers. His main research interests include speech signal processing and pattern recognition applications. He is an Associate Editor for IEEE/ACM TASLP and now is a member of IEEE SLTC and AASP-TC. He was the recipient of the 2018 IEEE Signal Processing Society Best Paper Award. His team won several champions of CHiME-4/CHiME-5/CHiME-6/CHiME-7 Challenge, SELD Task of 2020/2022/2023 DCASE Challenge, and DIHARD-III Challenge.
\end{IEEEbiography}

\vspace{0.5cm}
\begin{IEEEbiography}[{\includegraphics[width=1in, height=1.25in, clip,keepaspectratio]{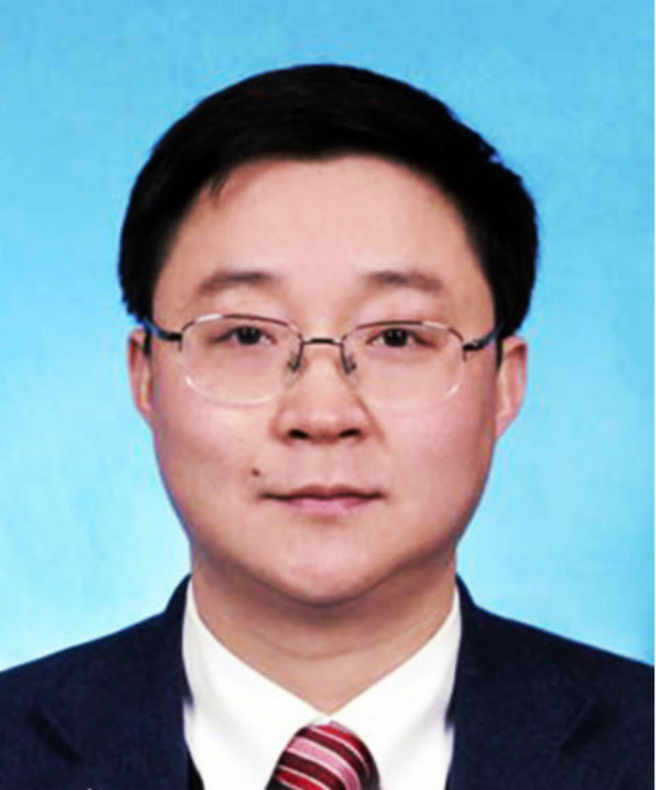}}]{Qing-Feng Liu}
received the B.Eng. and Ph.D., from the Department of Electronic Engineering and Information Science, University of Science and Technology of China (USTC), Hefei, China, in 1998 and 2003, respectively. He is the Founder, CEO, and President of iFLYTEK, the Director of the National Speech and Language Engineering Laboratory of China, a Professor and Doctoral Advisor with USTC, the Director General of the Union of Speech Industry of China, and the Director General of the Union of National University Student Innovation and Entrepreneurship.
\end{IEEEbiography}

\vspace{0.5cm}
\begin{IEEEbiography}
[{\includegraphics[width=1in, height=1.25in, clip,keepaspectratio]{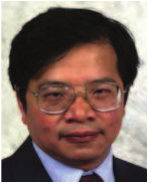}}]{Chin-Hui Lee}
(Fellow, IEEE) is a Professor with the School of Electrical and Computer Engineering, Georgia Institute of Technology. Before joining academia in 2001, he had 20 years of industrial experience, ending at Bell Laboratories, Murray Hill, NJ, USA, as a Distinguished Member of Technical Staff, and the Director of the Dialogue Systems Research Department. He has authored or coauthored more than 600 papers and patents. They have been more than 55,000 times with an $h$-index of 80 on Google Scholar. He has received numerous awards, including the Bell Labs President's Gold Award in 1998. He also won the 2006 Technical Achievement Award from IEEE Signal Processing Society for Exceptional Contributions to the Field of Automatic Speech Recognition. In 2012, he was invited by ICASSP to give a plenary talk on the future of speech recognition. In the same year, he was awarded the ISCA Medal for Scientific Achievement for "Pioneering and Seminal Contributions to the Principles and Practices of Automatic Speech and Speaker Recognition".
\end{IEEEbiography}

\end{document}